\documentclass[conference]{IEEEtran}
\IEEEoverridecommandlockouts
\usepackage{cite}
\usepackage{amsmath,amssymb,amsfonts}
\usepackage{graphicx}
\usepackage{textcomp}
\usepackage{xcolor}
\usepackage{amsmath}      % For advanced math typesetting
\usepackage{graphicx}     % For including images
\usepackage{amsmath}    % For advanced math typesetting
\usepackage{amssymb}    % For additional math symbols

\usepackage{algorithm}
\usepackage{algpseudocode}
\usepackage{xcolor}

\usepackage{amsmath}    % For mathematical expressions
\usepackage{graphicx}   % For including graphics

\def\BibTeX{{\rm B\kern-.05em{\sc i\kern-.025em b}\kern-.08em
    T\kern-.1667em\lower.7ex\hbox{E}\kern-.125emX}}
\begin{document}

\title{\textit{DemoCraft}: Using In-Context Learning to Improve Code Generation in Large Language Models\\
}

\author{\IEEEauthorblockN{Kapu Nirmal Joshua}
\IEEEauthorblockA{\textit{Department of Electrical Engineering} \\
\textit{Indian Institute of Technology Kanpur}\\
nirmalj21@iitk.ac.in}
\and
\IEEEauthorblockN{Mihit Sreejith}
\IEEEauthorblockA{\textit{Department of Computer Science and Engineering}}
\textit{Indian Institute of Technology Guwahati}\\
s.mihit@iitg.ac.in}

\maketitle

\begin{abstract}
Generating executable code from natural language instructions using Large Language Models (LLMs) poses challenges such as semantic ambiguity and understanding task-specific contexts. To address these issues, we propose a system called \textit{DemoCraft}, which enhances code generation by leveraging in-context learning and demonstration selection, combined with latent concept learning. Latent concept learning introduces additional concept tokens, which are trainable embeddings that capture task-specific knowledge. We then test our system on two major datasets: \textit{MBPP} and \textit{Humaneval}. Our experimental results demonstrate that the proposed system achieves an approximate \textit{2x} increase in the \textit{pass@k} metric compared to baseline models. Furthermore, we introduce two novel evaluation metrics: \textit{correctness@k} and \textit{similarity@k}. Our empirical studies indicate that our system attains nearly a \textit{3x} improvement in these metrics as well.
\end{abstract}


\section{Introduction}
The problem of generating code from natural language using Large Language Models (LLMs) involves creating systems capable of translating human language instructions into executable code accurately. This requires the LLM to understand the semantics of the natural language input, grasp the intent behind the instructions, and convert it into syntactically correct and functional code in a specified programming language. Key challenges include handling ambiguous or imprecise language, ensuring the generated code is both correct and efficient, and covering a wide range of programming scenarios and languages.

\begin{figure}[htbp]
\centering
\includegraphics[width=1\linewidth]{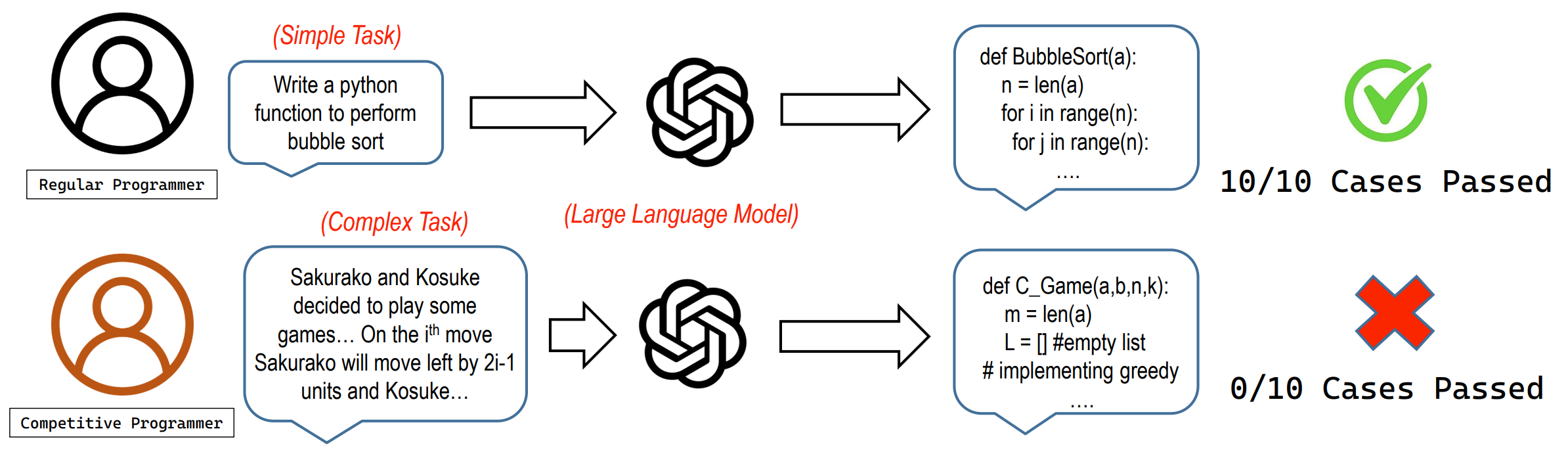}
\caption{Large Language Models struggling at Code Generation}
\label{fig}
\end{figure}

Code generation remains a significant challenge for large language models, as evidenced by Google's \textit{AlphaCode}\cite{b1}, developed specifically for competitive programming tasks. When evaluated on the \textit{CodeContests} benchmark, AlphaCode achieves a maximum Codeforces rating of only \textbf{1238}, placing it in approximately the top 28th percentile. Furthermore, a comprehensive survey on code generation using large language models~\cite{b2} reports a maximum \textit{pass@1} rate of around 30\%. These studies have been conducted under zero-shot conditions, highlighting the necessity for few-shot learning approaches. Few-shot learning allows models to leverage relevant demonstrations associated with the prompt prior to generating the output, potentially improving performance.
\section{Problem Behind Selecting Demonstrations}

In-context learning operates by pre-pending a series of demonstrations—examples of prompts and corresponding answers—before the final prompt that the model needs to solve. This setup effectively guides the model, allowing it to leverage patterns from prior examples to generate improved responses. By selecting demonstrations that closely match the problem at hand, we can significantly enhance the model's performance on complex tasks like code generation.

\begin{figure}[htbp]
\centering
\includegraphics[width=1\linewidth]{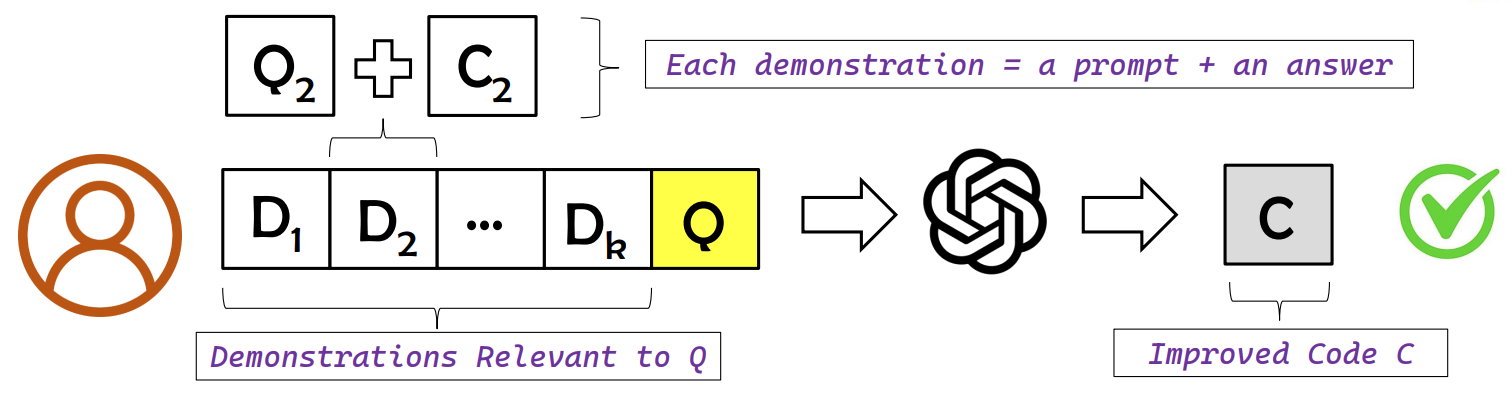}
\caption{Few Shot Learning Pipeline}
\label{fig}
\end{figure}

However, selecting relevant demonstrations is a challenging task in itself. Semantic similarity-based selection, a commonly used approach, attempts to identify demonstrations that share high textual similarity with the prompt. While this method may capture surface-level relationships, it often fails to consider the deeper task requirements.

For instance, in competitive programming contexts like Codeforces, problem statements frequently involve recurring character names like "Alice" and "Bob," often engaging in a hypothetical game. A semantic similarity-based approach might assume that any problem mentioning "Alice and Bob playing a game" is contextually relevant to another problem with similar phrasing. However, while these problems may seem alike in language, they can differ significantly in their underlying algorithms. One "Alice and Bob" problem may require a dynamic programming approach, while another could involve graph theory or combinatorial analysis. As a result, semantically similar demonstrations might mislead the model, offering examples that match the language but fail to provide the right procedural insights.

This is where our system, \textit{DemoCraft}, becomes instrumental. \textit{DemoCraft} utilizes a latent concept-based selection algorithm to analyze and select demonstrations that are aligned not only in linguistic features but also in conceptual depth. By focusing on the intrinsic structure of computational problems, \textit{DemoCraft} identifies demonstrations that share the same reasoning paradigms or algorithmic strategies necessary to solve the target prompt. For instance, when presented with a complex binary search or dynamic programming problem, \textit{DemoCraft} is capable of prioritizing demonstrations that involve these specific techniques over those with mere superficial similarity, thereby ensuring that the model is provided with the most contextually relevant guidance.

\begin{figure}[htbp]
\centering
\includegraphics[width=1\linewidth]{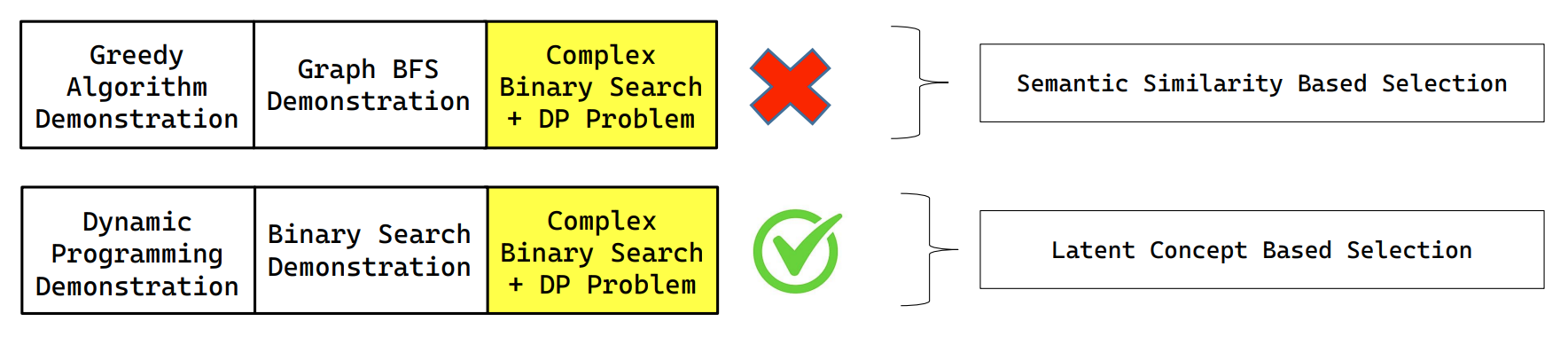}
\caption{Demonstration Selection with Latent Concept Learning}
\label{fig}
\end{figure}

\section{DemoCraft: System Details}
In this section, we provide a detailed technical description of our system architecture, which consists of three primary components: the Latent Concept Learning module, the Task Concept Probability Calculation module, and the Demonstration Selector.

\subsection{Latent Concept Learning}
In this stage, we introduce additional tokens \cite{b6}, referred to as \textit{concept tokens}, to enable the model to learn task-specific features for a given task. These concept tokens function as specialized units within the language model, representing knowledge specific to the task. Incorporating these tokens allows the model to predict the structure and requirements of the task more effectively.

We aim to find the optimal value of the variable $\theta_d$ for each task $d$ in the set of tasks $T$. The variable $\theta_d$, referred to as the \textit{latent concept variable}, is intended to capture the essential characteristics of each task to maximize the model's predictive accuracy. Mathematically, the optimal $\theta_d$ maximizes the probability of the correct output given the input, achieved through the Bayes optimal classifier defined as
\begin{equation}
\theta_d = \arg\max_{\theta_d} P_M^d(Y \mid \theta_d, X)
\label{eq:bayes_optimal}
\end{equation}
where $P_M^d(Y \mid \theta_d, X)$ is the probability that the model $M$ assigns to the output $Y$ given the input $X$ and task-specific variable $\theta_d$.

To train the model to make better predictions, we aim to find $\theta_d$ that minimizes the cross-entropy loss. This involves minimizing the negative expected log probability:
\begin{equation}
\hat\theta_d = \arg \min_{\theta_d} -\mathbb{E}_{X,Y,d} \left[ \log P_M^d(Y \mid \theta_d, X) \right]
\label{eq:cross_entropy}
\end{equation}

We align $\hat\theta_d$ with the token embedding space by introducing new tokens—our concept tokens—into the model's vocabulary. These tokens represent the task concept $\theta_d$, allowing the model to utilize them within its regular vocabulary. Following methods proposed by Lester et al.~\cite{b3}, we add $c$ new concept tokens, denoted as $\hat{\theta}_d$, to represent each task's concept. The embeddings of these new tokens, $E_{\text{new}}(\hat{\theta}_d)$, are fine-tuned specifically for the task while keeping the rest of the language model's parameters frozen. This approach enables the model to focus on learning the nuances of $\theta_d$ without altering its general language capabilities. The parameter $c$, representing the number of concept tokens, is treated as a hyperparameter adjustable based on task requirements.

During training, the $c$ concept tokens associated with $\hat\theta_d$ are prepended to the input $X$ (or output $Y$) to condition the model on the specific task, providing task-specific context that enhances predictive performance.

\begin{figure}[htbp]
\centering
\includegraphics[width=1\linewidth]{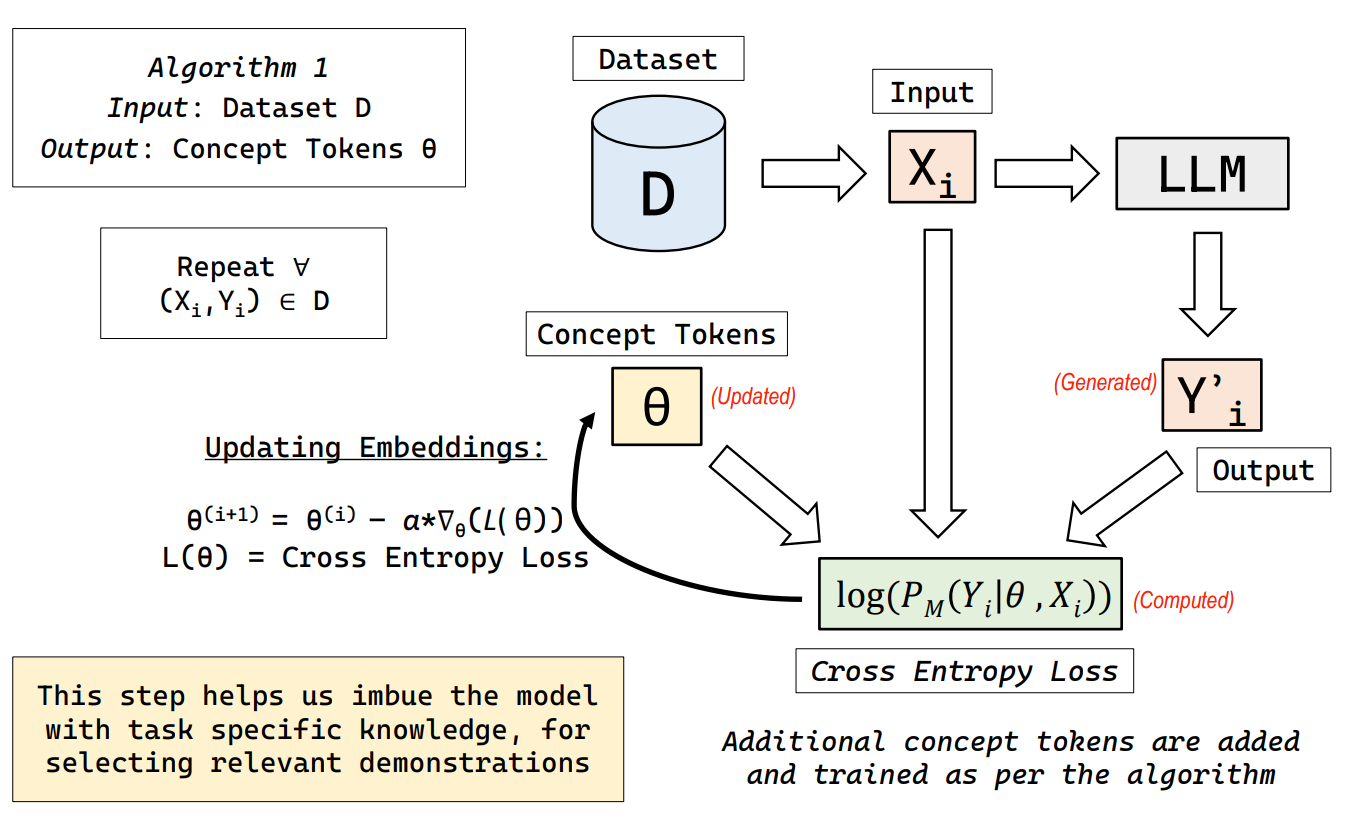}
\caption{Latent Concept Learning Module}
\label{fig:latent_concept_learning}
\end{figure}

This process is illustrated in Figure~\ref{fig:latent_concept_learning}, which provides a flowchart for the latent concept learning method. The flow depicts how, starting from a dataset $D$, the input $X_i$ is fed into the model along with the updated concept tokens $\hat\theta$. The model generates the output $Y_i'$, and the cross-entropy loss $\log P_M(Y_i \mid \theta, X_i)$ is computed to update $\theta$. This iterative training process enables the model to understand and adapt to the task-specific requirements embedded in $\theta$, leading to more relevant demonstration selections in \textit{DemoCraft}.

\subsection{Task Concept Probability Calculation}
In the \textit{Task Concept Probability Calculation} stage, our objective is to quantify how well each demonstration aligns with the target task. This involves calculating the relevance of each input-output pair $(X_i, Y_i)$ within the context of the task's specific requirements.

Leveraging the previously trained concept tokens $\theta$, we evaluate the suitability of input-output pairs from our dataset $\mathcal{D}$. For each pair $(X_i, Y_i)$, we compute the probability $P_M(\theta \mid Y_i, X_i)$, which measures the degree to which the demonstration aligns with the task-specific concept encapsulated by $\theta$. This probability serves as an evaluative metric, where higher values indicate stronger alignment with the task.

Formally, the task concept probability is calculated using Bayes' theorem:
\begin{equation}
P_M(\theta \mid Y_i, X_i) = \frac{P_M(Y_i, X_i \mid \theta) P_M(\theta)}{P_M(Y_i, X_i)},
\label{eq:task_concept_probability}
\end{equation}
where:
\begin{itemize}
    \item $P_M(\theta \mid Y_i, X_i)$ is the posterior probability of the concept tokens given the demonstration pair.
    \item $P_M(Y_i, X_i \mid \theta)$ is the likelihood of the demonstration pair given the concept tokens.
    \item $P_M(\theta)$ is the prior probability of the concept tokens.
    \item $P_M(Y_i, X_i)$ is the marginal probability of the demonstration pair.
\end{itemize}

In this stage, the large language model $M$ operates in an evaluative capacity; it computes the task concept probabilities based on its learned representations without undergoing further fine-tuning. By assigning task concept probabilities to each demonstration, we gain insights into their relative relevance, which is crucial for selecting the most appropriate demonstrations in subsequent stages.

\begin{figure}[htbp]
\centering
\includegraphics[width=1\linewidth]{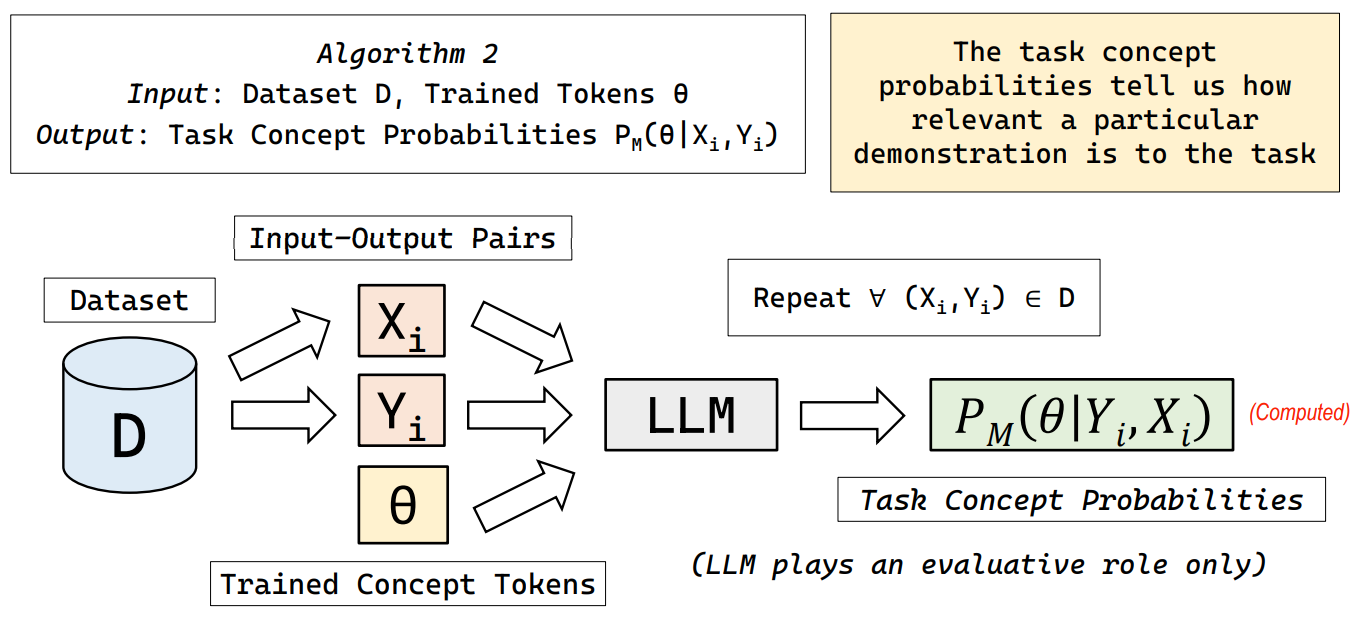}
\caption{Task Concept Probability Calculation Module}
\label{fig:task_concept_probability_calculation}
\end{figure}

This process is illustrated in Figure~\ref{fig:task_concept_probability_calculation}, which outlines how input-output pairs, along with the trained concept tokens $\theta$, are processed through the model to compute the task concept probabilities $P_M(\theta \mid Y_i, X_i)$ for each pair $(X_i, Y_i)$.

\subsection{Demonstration Selection}
In the \textit{Demonstration Selection} stage, our objective is to identify the most relevant demonstrations for a given task prompt. Having computed the task concept probability $P_M(\theta \mid Y_i, X_i)$ for each demonstration pair $(X_i, Y_i)$ in our dataset $\mathcal{D}$, we proceed to select the top $k$ demonstrations that align most closely with the task-specific concept $\theta$.

\begin{figure}[htbp]
\centering
\includegraphics[width=1\linewidth]{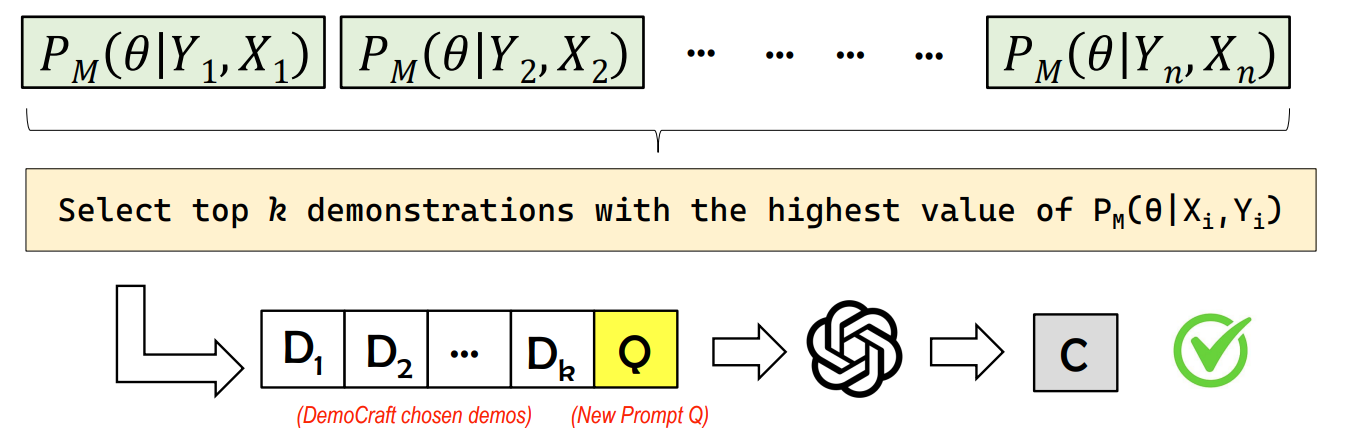}
\caption{Demonstration Selection Module}
\label{fig:demonstration_selection}
\end{figure}

We rank all demonstration pairs based on their computed task concept probabilities and select the top $k$ pairs with the highest values of $P_M(\theta \mid Y_i, X_i)$. This selection process ensures that we retain demonstrations that are most contextually relevant to the task at hand. By focusing on the highest probability values, we choose examples that the model has identified as highly aligned with the desired task-specific features. This maximizes the likelihood that these demonstrations will enhance the model's understanding and performance when generating responses for the target prompt.

This process is illustrated in Figure~\ref{fig:demonstration_selection}, which shows how we systematically select the top $k$ demonstrations with the highest alignment scores, ultimately constructing a refined set of examples tailored to optimize the model's responses for the given prompt.

\subsection{Final System Diagram}

\textit{DemoCraft} extends the foundational concepts discussed—namely, latent concept learning and task concept probability calculation—to operate across multiple datasets. This enables the model to learn a comprehensive set of concept tokens, each corresponding to distinct task types denoted by $\theta_1, \theta_2, \ldots, \theta_k$. Once trained, these concept tokens allow the system to retrieve relevant demonstrations from a diverse range of sources.

When a new prompt $Q$ is provided, \textit{DemoCraft} evaluates it by calculating probabilities over both the learned concept tokens and potential demonstration pairs $(X_j, Y_j)$ from the dataset $\mathcal{D}$. This involves a two-step process:

\begin{enumerate}
    \item For each concept token $\theta_i$, compute the probability $P_M(\theta_i \mid X_j, Y_j)$ for all demonstration pairs $(X_j, Y_j) \in \mathcal{D}$.
    \item Maximize this probability over both $\theta_i$ and $(X_j, Y_j)$ to select the top $k$ demonstrations:
    \begin{equation}
    \label{eq:demo_selection}
    \{(X_{i^*}, Y_{i^*})\} = \arg \max_{\theta_i, (X_j, Y_j)} P_M(\theta_i \mid X_j, Y_j),
    \end{equation}
\end{enumerate}
where $\{(X_{i^*}, Y_{i^*})\}$ denotes the set of top $k$ demonstrations that best align with the task-specific requirements of $Q$. This approach leverages both the learned task-specific knowledge encapsulated in the concept tokens and the diversity of the dataset, ensuring a refined and targeted selection process.

\begin{figure}[htbp]
\centering
\includegraphics[width=1\linewidth]{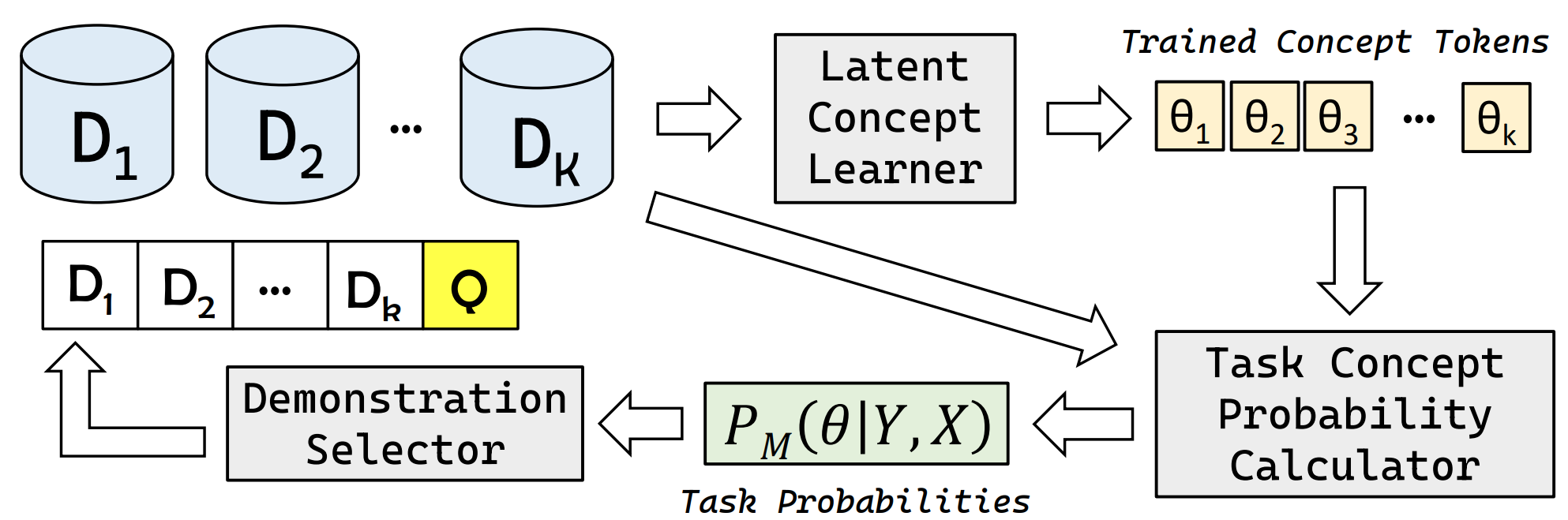}
\caption{\textit{DemoCraft} System Flowchart}
\label{fig:democraft_system}
\end{figure}

The overall system flowchart is provided in Figure~\ref{fig:democraft_system}, illustrating how the trained concept tokens, task probability calculator, and demonstration selector operate in unison to choose the most relevant examples for each new prompt.

\section{Experiments}

In this section, we highlight our experimental metrics and the conditions under which we conducted the experiments.

\subsection{Evaluation Metrics}

We evaluate our model using three primary metrics:

\begin{enumerate}
    \item \textit{\textbf{pass@k}}: This metric measures the probability that at least one of the top $k$ generated code samples passes all the test cases for a given problem. Suppose for each problem we generate $n$ code samples, out of which $c$ samples are correct (i.e., they pass all the unit tests). The pass@k is calculated as:
    \begin{equation}
    \text{pass@}k = \mathbb{E}_{D} \left[ 1 - \frac{\binom{n - c}{k}}{\binom{n}{k}} \right],
    \label{eq:passk}
    \end{equation}
    where $\mathbb{E}_{D}$ denotes the expectation over the dataset $D$, and $\binom{n}{k}$ is the binomial coefficient representing the number of ways to choose $k$ samples out of $n$.

    \item \textit{\textbf{correctness@k}}: This metric is defined as the average precision of the model over the entire dataset when $k$ outputs are generated per prompt. For each prompt, if the model generates $k$ outputs and $c$ of them are correct, the correctness for that prompt is calculated as:
    \begin{equation}
    \text{correctness@}k = \mathbb{E}_{D} \left[ \frac{c}{k} \right],
    \label{eq:correctnessk}
    \end{equation}
    where $\mathbb{E}_{D}$ denotes the expectation over the dataset $D$.

    \item \textit{\textbf{similarity@k}}: This metric measures the average similarity between the working codes generated by the model and the golden solution provided in the dataset. For each prompt, let $S$ be the set of all generated codes that pass all the test cases (i.e., working codes), and let $y$ be the golden solution from the dataset. The similarity@k is defined as:
    \begin{equation}
    \text{similarity@}k = \mathbb{E}_{D} \left[ \frac{1}{|S|} \sum_{y_i \in S} \text{sim}(y_i, y) \right],
    \label{eq:similarityk}
    \end{equation}
    where $\text{sim}(y_i, y)$ is a similarity function between the generated code $y_i$ and the golden solution $y$, and $|S|$ is the number of working codes for that prompt. The outer expectation $\mathbb{E}_{D}$ is taken over all prompts in the dataset $D$. The similarity function used over here is the \textit{edit distance} metric, provided in the standard \textit{NLTK} library.
\end{enumerate}

\subsection{Datasets and Models}

We conducted our experiments using the following datasets and model:

\begin{enumerate}
    \item \textbf{MBPP}: The \textit{Mostly Basic Python Problems} (MBPP) dataset \cite{b4} consists of 427 programming problems designed for code generation tasks. Each problem includes a natural language description, the corresponding code solution, and three unit tests. The programming language used is Python.

    \item \textbf{HumanEval}: The HumanEval dataset \cite{b5} comprises 164 programming tasks focused on code completion. Each task provides a function signature and a docstring describing the desired functionality. The solutions are written in C++, and each problem includes approximately seven unit tests, making it a stricter benchmark than MBPP.
\end{enumerate}

Due to resource constraints, we evaluated the performance of our system using the \textit{SantaCoder} model. SantaCoder is a transformer-based language model with 1.1 billion parameters, pretrained on a large corpus of code in multiple programming languages, including Python and C++. It is designed to generate syntactically correct and functionally meaningful code snippets. We conducted our experiments using Google Colab's T4 GPU, which provided sufficient computational resources for our evaluations without compromising performance.

\subsection{Baselines}

We compare our system against the following baseline methods:

\begin{enumerate}
    \item \textbf{Semantic Selection}: In this baseline, we select demonstrations from the dataset purely based on their semantic similarity to the given prompt $x$. Let the dataset be $\mathcal{D} = \{(x_i, y_i)\}_{i=1}^{n}$, where $x_i$ are the prompts and $y_i$ are the corresponding outputs. For each $x_i$ in the dataset, we compute the similarity score $\text{sim}(x, x_i)$ between the given prompt $x$ and each dataset prompt $x_i$. We then select the top $k$ demonstrations with the highest similarity scores:
    \begin{equation}
    \label{eq:semantic_similarity}
    \{(x_{i^*}, y_{i^*})\} = \arg \max_{\substack{(x_i, y_i) \in \mathcal{D} \\ i = 1, \ldots, n}} \text{sim}(x, x_i),
    \end{equation}
    where $\{(x_{i^*}, y_{i^*})\}$ denotes the set of top $k$ demonstrations selected. The \textit{sim(.)} function used here is the standard edit distance, implemented using the \textit{NLTK} library.

    \item \textbf{Random Selection}: In this baseline, we randomly select $k$ demonstrations from the dataset $\mathcal{D}$ without considering their relevance to the given prompt $x$. This method serves as a control to evaluate the impact of demonstration selection strategies on the model's performance.
\end{enumerate}

\section{Results}
In this section, we present the results of our experiments on both the \textit{MBPP} and \textit{HumanEval} datasets. Table 1 shows the results for the \textit{MBPP} dataset, while Table 2 presents the results for the \textit{HumanEval} dataset.
\begin{table}[htbp]
\caption{Evaluation Results on MBPP}
\begin{center}
\begin{tabular}{|c|c|c|c|}
\hline
\textbf{Parameter} & \textbf{Semantic} & \textbf{\textit{DemoCraft}} & \textbf{Random} \\
\hline
\textit{correctness@5} & 2\% & 7.2\% & 1.5\% \\
\hline
\textit{correctness@20} & 0.5\% & 6.0\% & 0.3\% \\
\hline
\textit{correctness@100} & 0.3\% & 5.0\% & 0.2\% \\
\hline
\textit{similarity@5} & 0.77\% & 3.0\% & 0.5\% \\
\hline
\textit{similarity@20} & 0.771\% & 3.5\% & 0.4\% \\
\hline
\textit{similarity@100} & 2.7\% & 7.0\% & 1.8\% \\
\hline
\textit{pass@1} & 0.6\% & 4.0\% & 0.2\% \\
\hline
\textit{pass@10} & 6.07\% & 11.5\% & 5.0\% \\
\hline
\textit{pass@100} & 20\% & \textbf{27.0}\% & 15.0\% \\
\hline
\end{tabular}
\label{tab:mbpp_evaluation_results}
\end{center}
\end{table}

\begin{table}[htbp]
\caption{Evaluation Results on HumanEval}
\begin{center}
\begin{tabular}{|c|c|c|c|}
\hline
\textbf{Parameter} & \textbf{Semantic} & \textbf{\textit{DemoCraft}} & \textbf{Random} \\
\hline
\textit{correctness@5} & 0.1\% & 1.2\% & 0.2\% \\
\hline
\textit{correctness@20} & 0.04\% & 1.1\% & 0.03\% \\
\hline
\textit{correctness@100} & 0.008\% & 1.0\% & 0.005\% \\
\hline
\textit{similarity@5} & 0.91\% & 3.5\% & 0.8\% \\
\hline
\textit{similarity@20} & 0.92\% & 4.0\% & 0.7\% \\
\hline
\textit{similarity@100} & 3\% & 7.5\% & 2\% \\
\hline
\textit{pass@1} & 0.3\% & 2.0\% & 0.4\% \\
\hline
\textit{pass@10} & 4.56\% & 8.0\% & 3\% \\
\hline
\textit{pass@100} & 13.2\% & 18.5\% & 10\% \\
\hline
\end{tabular}
\label{tab:humaneval_evaluation_results}
\end{center}
\end{table}

\begin{figure*}[htbp]
    \centering
    \begin{minipage}[b]{0.45\textwidth}
        \centering
        \includegraphics[width=\linewidth]{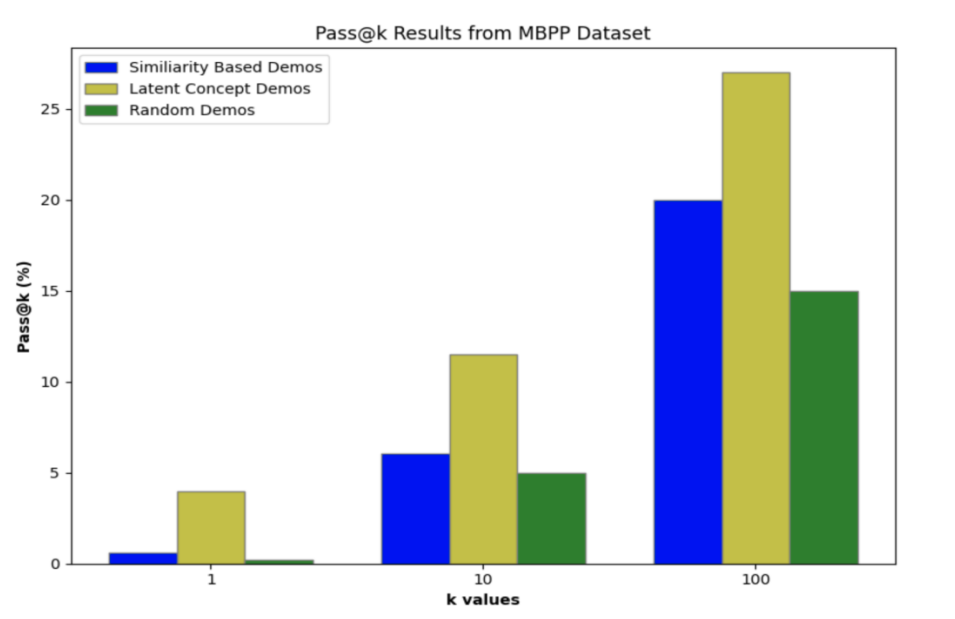}
        \caption{}
        \label{fig:passk_mbpp}
    \end{minipage}
    \hspace{0.10\textwidth} % Adjust the horizontal space as needed
    \begin{minipage}[b]{0.40\textwidth}
        \centering
        \includegraphics[width=\linewidth]{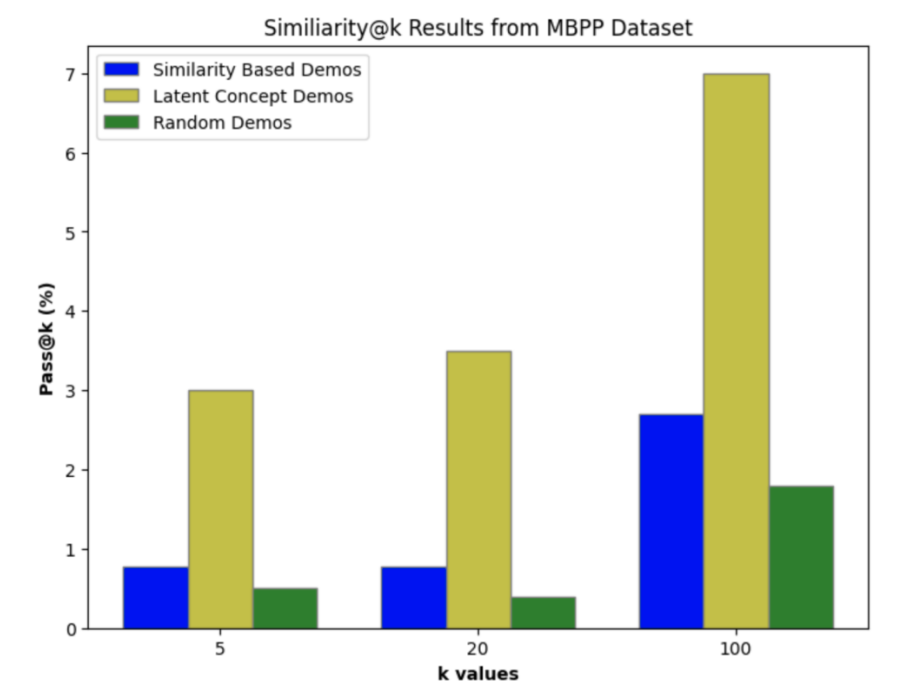}
        \caption{}
        \label{fig:similarityk_mbpp}
    \end{minipage}
    \hspace{0.10\textwidth} % Adjust the horizontal space as needed
    \begin{minipage}[b]{0.45\textwidth}
        \centering
        \includegraphics[width=\linewidth]{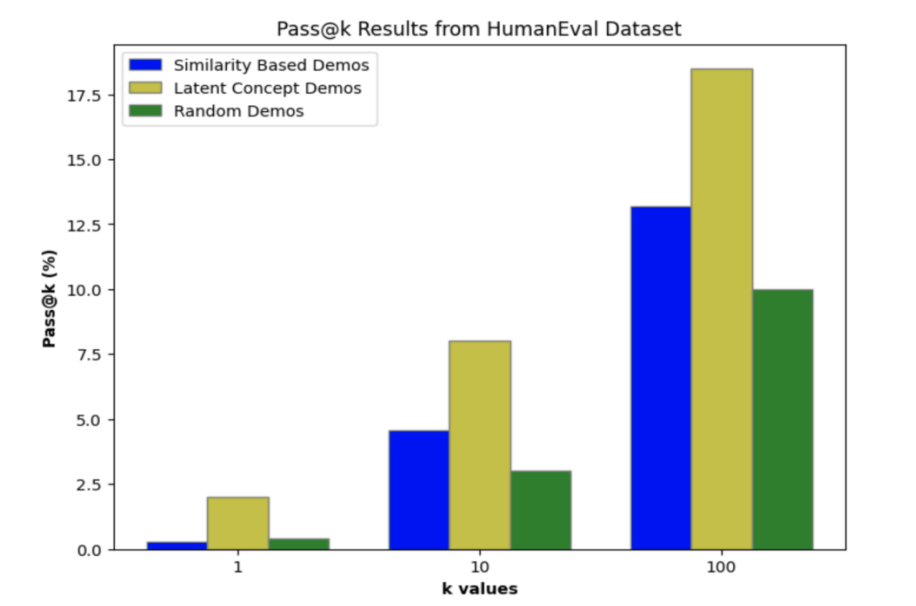}
        \caption{}
        \label{fig:similarityk_mbpp}
    \end{minipage}
    \hspace{0.10\textwidth} % Adjust the horizontal space as needed
    \begin{minipage}[b]{0.40\textwidth}
        \centering
        \includegraphics[width=\linewidth]{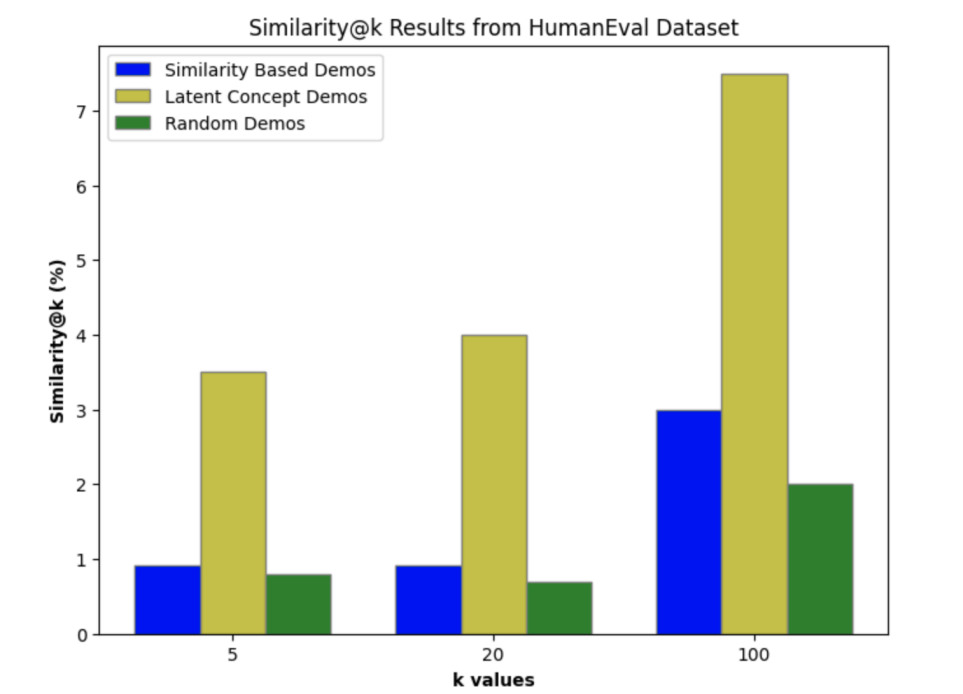}
        \caption{}
        \label{fig:similarityk_mbpp}
    \end{minipage}
\end{figure*}
The results show that demonstrations chosen by \textit{DemoCraft} consistently outperform other selection methods. This superiority arises from \textit{DemoCraft}'s encoding of task-specific knowledge through specialized token embeddings tailored to each task. 

\section{Conclusion}

In this paper, we presented \textit{DemoCraft}, a demonstration selection framework that enhances code generation models by leveraging task-specific knowledge through latent concept learning. \textit{DemoCraft} introduces specialized token embeddings tailored to each task, enabling the model to internalize underlying concepts effectively. Our evaluations on the \textit{MBPP} and \textit{HumanEval} datasets, utilizing the metrics \textit{pass@k}, \textit{correctness@k}, and \textit{similarity@k}, demonstrate that \textit{DemoCraft} consistently outperforms baseline methods, including semantic similarity-based and random selection approaches. These results highlight the efficacy of targeted demonstration selection in improving code generation accuracy and functionality. Future work will explore the integration of \textit{DemoCraft} with larger language models and its application to diverse domains, including software engineering and competitive programming.

\section*{Acknowledgements}

We acknowledge Dr. Amar Prakash Azad and Dr. Brij Kumar Chavda from IBM Research Bangalore for their invaluable support and mentorship, which were instrumental to the success of this project.

\vspace{12pt}

\end{document}